# Non-Cooperative MDS Coding Game In CSMA Wireless Networks

Mohamed Lamine BOUCENNA

Electromagnetism and Telecommunication laboratory
LET, University of Constantine1
Constantine, Algeria
boucenna.m.lamine@gmail.com

Malek BENSLAMA

Electromagnetism and Telecommunication laboratory
LET, University of Constantine1
Constantine, Algeria
Ma_benslama@yahoo.fr

*Abstract*—Maximum distance separable erasure coding has been introduced in wireless networks based on random medium access protocols in order to recover collided and erased packets. So, this help to avoid retransmission process which weaken the network throughput and prolong the overall propagation delay. However, erasure coding may degrade the network performances by causing overload traffic due to added redundancy (parity) and also when extending the propagation delay due to additional calculations and encoding/decoding process time which can lead quickly the network into saturation. Here, the redundancy should be applied properly. In addition, multiple access make conflicting situations due to simultaneous access and selfish behaviour of users which affect too the network performances. For this purpose, we present a game model based on the use of game theory to analyse and manage multiple users that assuming random access protocol and erasure coding algorithm. Our model is used to operate the network at equilibrium by selecting a suitable type of erasure coding. Making the network work at equilibrium, leads to decrease collisions, decrease also the overall transmission delay and as well as increasing the throughput.

## I. INTRODUCTION

In wireless networks, we can distinguish different conflicting situations; as in crowded places like in city centers and parking spaces, where we found a lot of users with small separate distances, here we can observe simultaneous transmissions and also the phenomenon of exposed nodes that create collisions. Another type of situations is when the user is traveling in isolated or rural places, where we can find few users with big separate distances, and here we can observe a phenomenon of hidden nodes that causes the collision.

The Non-Persistent Carrier Sense Multiple Access (N-p CSMA) protocol adds important functionality to the random protocols and, as its name suggests, CSMA/CA (with collision avoidance) senses the medium to determine whether any other node (user) is transmitting. Despite all these preventative techniques, networks still suffer from packet collisions that reduce throughput and affecting system performances. So, we find recently many researches as in [1, 2] that propose different solutions to improve throughput and other performances of Wireless networks based on CSMA MAC protocols. Thus it is still difficult to fixe completely collisions under the multiple access protocols, which encourages and prompts researchers to work more to solve this issue.

In attempt to improve the efficiency of these protocols; several models and solutions have been proposed. When attempting to access medium or selecting suitable roots; users are selfish and each one want to success its transmissions in a short delay. Creating thus conflicts between users. In such situations and as it contains conflicting situations, CSMA networks are analyzed by Game theory. Authors in papers [3, 4] presented game theoretical analyze to find optimal performance of CSMA at Nash equilibrium basing on Markov chain model. Also a game study is presented in [5] to optimize energy consumption in wireless networks. A recent study of 1-persistent CSMA with variable collision length is presented in [6] to improve the network access performances during multimedia video transmission. Authors in [7] propose a new game model in MAC protocols based on the control of the back off interval of each user during his transmission in order to decrease the collusion rate and improve the throughput.

In the present paper, we study the wireless network composed by at least two active users that communicate between them or with a remote station. We treat active users as actives nodes, and we try to improve the network quality via decreasing collusion rate and packet transmission delay. We treat the network at the access method level; here we have a random access method based on Non-Persistent CSMA (N-P CSMA) protocol. In the next section we review the relevant literature by introducing the Maximum Distance

Separable erasure coding (MDS EC) scheme and the advantages that can import to the network. We introduce then, the Game theory method and its importance for wireless networks. In section III, we present our model and we express the utility function. Finally, we investigate the network equilibrium existence to analyse the overall packet transmission delay.

## II. MDS EC SCHEME

EC is an error correction technique involved from Reed-Solomon error correcting codes. As in RS algorithm, EC (N,k) encodes k original packets with m redundant packets to N codded packets as a code word. So, the encoding (decoding) rate is k/N, with a minimum distance $d_{min} = N - k + 1$, when theoretically; a code with $d_{min}$ has no information lost even with $d_{min} - 1$ erasures. When errors (e) = m, such an erasure code is called the Maximum Distance Separable or simply MDS code. An MDS erasure code is optimal in terms of space efficiency for a designed erasure recovery degree (e) and, therefore, is most desired in many systems and applications, including data storage systems [8]. RS codes have the property of Maximum Separable Distance, which means that any (N) out of (N + k) is sufficient to rebuild the original data block, thus, RS codes can tolerate (k) simultaneous failures. RS codes present a trade-off between higher fault tolerance and lower storage overhead depending on the parameters (n) and (k) [9]. So, EC presents high reliability in recovering original packets at the reception by recovering collided and also erased packets.

In MDS EC, we send *N* packets instead of *k* original packets. We create thus (*N-k*) redundant packets as shown in Fig. 1. Generating redundant packets is following certain functions in coding process that is deeply described in literature [10] and should respecting the code rate *R=k/N*. At the reception, we can recover the *G* original packets if we receive successfully *N'* packets out of *N* codded packets.

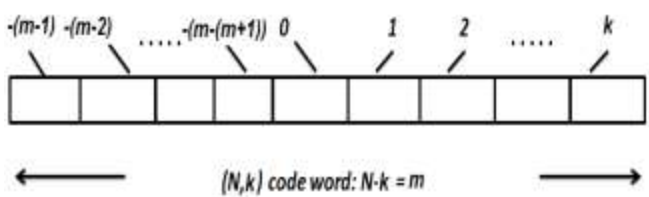

Figure 1. Code word form in *MDS EC* with *m* redundant packets.

If the redundancy is well structured (selected), we could achieve high transmission efficiency without any recourse to retransmission that extends the transmission delay and weakens the network throughput.

## III. GAME THEORY

Game theory is a discipline aiming to model situations in which decision makers have to make specific actions that have mutual-possibly conflicting-consequences [11]. It has been used primarily in economics, in order to model competition between companies. We distinguish two types of games: in non-cooperative games, each player selects strategies without coordination with others. On the other hand, in a cooperative game, the players cooperatively try to come to an agreement, and the players have a choice to deal with each other so that they can gain maximum benefit, which is higher than what they could have obtained by playing the game without cooperation [12]. Once such a game has been defined, game theory defines a solution concept which attempts to specify what we should expect to occur if rational players play the game. The most widely known solution concept is the Nash Equilibrium.

The Nash equilibrium is a solution concept of a game involving two or more players, in which no player has anything to gain by changing only his own strategy unilaterally. If each player has chosen a strategy and no player can benefit by changing his strategy while the other players keep their unchanged, then the current set of strategy choices, and the corresponding payoffs constitute a Nash equilibrium (NE) [13].

The use of game theory in wireless networks is becoming very large due to the benefic that can import. Many recent researches have applied game theory and proved its efficiency in such a domain as it may be used to study the cooperation (or noncooperation) among nodes or terminals forming the network. For wireless networks based on random access protocols, we find two good reasons that making game theory suitable tool for analyzing distributed medium access. First, the contention-based nature of the medium access presents a natural application domain for Non-cooperative game theory. Secondly, it is possible to conceive of selfish users in future choosing their individual access strategies to optimize their own selfish utilities [8].

## IV. PROBLEM AND CONFLICTING SITUATIONS

MDS EC is employed to increase the network reliability by recovering collided and erased packets during the transmission. However, introducing the redundancy could increase the traffic load and may disturb the network. So, the redundancy develops the traffic overload and could lead the network to its saturation as explained in below figure. Fig. 3 gives an example of three active nodes composing the CSMA network when each one uses a different type of redundancy. We can understand how redundancy could disturb the network with presence of other undesired factors like exposed and hidden nodes. Then, with big backlogged overload traffic, the network reaches quickly its saturation mode, and no one can then transmit. For that, selecting a suitable redundancy is a key factor.

In addition, when users in CSMA network transmit simultaneously, though, all access fail. Also, each user tries to accomplish its transmission successfully in the shortest delay. Such behaviours create conflicting situations among user. Here lies the importance of the game theory that gives us a good determination of such situations.

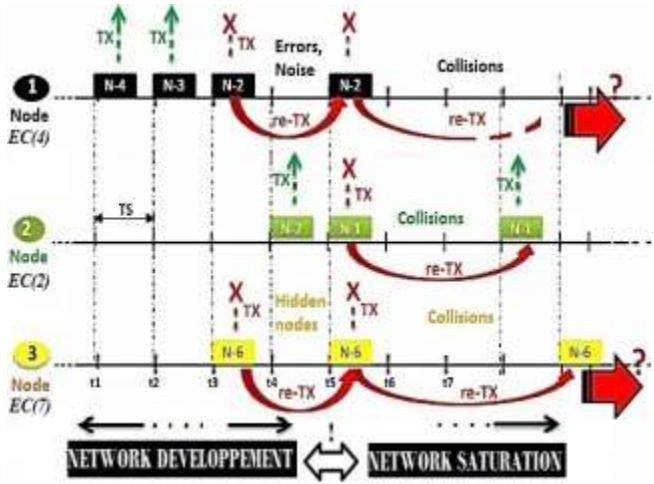

Figure 2. Errors, hidden nodes and collisions that leads to saturation in CSMA networks.

## V. FORMULATION OF THE GAME MODEL

The network model is composed of $M$ nodes that are willing to transmit packets. To access the canal, users assume N-P CSMA MAC protocol. Time is devised into equal periods called time slots (TS) with length equal to the transmission time of the single packet.

### A. Players

Players in our model are the nodes that enter the game when they start transmission (active nodes). We denote the player by $i$. $i = 1,...,M$. As in reality, players do not practice any cooperation between them and they are free and selfish. We go now to identify our game components and characteristics like players' strategies and profiles, the utility of each player. Then we investigate the game equilibrium existence and we study the game convergence and its behavior regarding the equilibrium strategy.

### B. Game strategies

The strategy of players is to select a number of redundant packets (type of redundancy) that allows to establish the MDS EC and also to satisfy the own utility of the player. The player then will be stable and wishes to continue on the same process. By the way, the player could all time changing its strategy in depending of the present network mode in order always to satisfy its utility. So, if the network is going to be saturated; the player prefers then to change its strategy by taking a small number of redundant packets to do not lose a big quantity of data and also to be able to send its data quickly at the first opportunity (free TS). Satisfying the utility means to maximize or to minimize the outcome of the player depending on the nature of the utility function. It means also to take the utility to its optimal value.

We denote the set of strategies for players by $S$, and the strategy of the player $i$ by $S_i$, and $S_{-i}$ denotes the strategies of all players other than player $i$. We can write then: $S = \{S_1, S_2, ..., S_M\} = \{r_1, r_2, ..., r_M\}$.

### C. The utility function

The utility function is defined as the outcome that can gain a player when applying a strategy. In our model we put the overall propagation delay as the utility amount that gains each player after playing his strategy. The overall propagation delay is defined as the average time from the start of sending packet until when it is successfully received. So, it may comprise the time of emission, retransmission and acknowledgment packets if necessary. Our objective is to minimize as possible the overall propagation delay and by consequently increasing the network throughput.

$u_i$ express the payoff utility assigned to player $i$. We can extract the propagation delay function from the expression of throughput in [14,15] that describes the impact of delay propagation (denoted by $D$) in CSMA wireless networks. So, we can write:

$$D = \frac{1}{qR}\left[\left(\frac{(1+M_d.R)Th}{M_d.R}\right)^{(1-M_d)} - 1\right]. \quad (1)$$

According to equation (2); we can observe the relationship between the throughput and the overall propagation delay (utility). The utility function for player $i$ could be expressed as;

$$u_i = \frac{1}{qR}\left[\left(\frac{(1+M_d.R)Th}{M_d.R}\right)^{(1-M_d)} - 1\right]. \quad (2)$$

When, $R$ describe the probing rate of nodes, $Th$ is indicating the network thoughput, $M_d$ is the number of active users and $q$ is a constant ($q = 1.53$). We simplify (2) as:

$$u_i = \frac{1}{qR}\left[\left(\frac{X}{M_d.R}\right)^{(1-M_d)} - 1\right]. \quad (3)$$

When $X = (1 + M_d.R)Th$.

The packets' arrivals are assuming Poisson process with mean $\lambda$ packets/second. We can then express the $X$ as follow:

$$X = (1 + M_d.R).\left(\frac{(1-e^{-\alpha G})^M.P_f.e^{-\alpha G(M-1)}}{1+\alpha-e^{-\alpha G}}\right). \quad (4)$$

When, $\alpha$ assumed as the time needed to sense the neighbor node ($\alpha = 0.2$) and $G$ is the traffic load of our network. To study $P_f$, we follow the below progress:

While the arrival of packets is following the Poisson process, so we can say that the packet can be successfully transmitted with the probability $P_p$:

$$P_p = (e^{-\alpha\lambda(1+m/k)})(1-e^{-\alpha\lambda(1+m/k)})^{M-1}. \quad (5)$$

Known that $\alpha(1+\frac{r}{G})$ is the traffic load after the coding process.

We can then receive successfully at least $k$ codded packets with the probability:

$$P_k = \sum_{i=k}^{N}\binom{N}{i}P_p^i Q_p^{N-i}. \quad (6)$$

Where; $Q_p = 1 - P_p$. Also, we can receive only $k$ codded packets ($m < k$) when $m$ out of $n$ are original packets, with probability:

$$P_m = \binom{k}{m}\binom{N-k}{n-m}P_p^i Q_p^{N-n}. \quad (7)$$

Now, we can express $P_f$ as the probability that an original packet is received or recovered successfully by:

$$P_f = P_k + \sum_{n=1}^{k-1}\sum_{m=1}^{n}(P_q + m/k)P_C. \quad (8)$$

We replace the expression of $P_f$ in (3) to get the final function of our utility expressed in (2).

*D. Existence analysis of the equilibrium value*

Bellow theorems will help us to discuss the equilibrium existence and its evaluation.

*1) Theorem (Fudenberg and Triole, 1991):* Debreu, Glicksberg and Fan (1952) have Consider a strategic-form game whose strategy spaces Si are nonempty compact convex subsets of a Euclidean space. If the payoff functions Ui are continuous in S, and quasi-concave in Si, there exists a pure strategy Nash equilibrium (NE) [16].

*2) Theorem (Fudenberg and Triole, 1991):* An equilibrium exists for every concave n-person game. To find the NE of the game we analyze the player's best response function (Osborne, 2004). The best response of player i is the number of redundant packets that maximizes the utility function [16].

*3) Definition (Jerzy Konorsky, 2007):* An action S is the best reply to S-i, if $u_i(s, s_{-i}) \geq u_i(s', s_{-i})$ for all $s' \in S$. Let BR(s-i) denote the set of best replies to S-i. An NE is an action profile $s = (s_i, s_{-i})$ in which $s_i \in BR(s_{-i})$ for all i=1...M [17].

Starting from the above theorems and definition, we will first investigate whether it exists the value of $r_i$ where the optimal *Th* can be acquired.

We remark that the utility vary with the parameters $N_d$, $R$ and *Th*. When only the *Th* is depending on our strategy $S$. Referring to equation (2), we can observe the relation between the throughput and the utility function *D*. The throughput is inversely related the transmission delay. If *Th* increases then *D* decreases and if *Th* decreases *D* increases. Now, we investigate whether there exist or not the value of $r=r^*$ which making the utility in its minimum value (ei: Max(*th*)).

Figure 4 plots the behavior of *Th* by varying $r$. It confirms that operating at $r = 0$ leads to network collapse. However there exists an optimal point of operation ($r = 5$) packets at which the throughput is maximized $Th(\rho = 5) = \arg\max Th$ and it corresponds to $D(r = 5) = \arg\min D$ for every node in the network. We refer to this optimal point of operation as $r^*$ witch corresponds to the Nash equilibrium for our scheme. It shows that when $r \leq r^*$, congestion dominates and results in higher propagation delay. Whereas, when $r \geq r^*$ the delay increase. Thus $D(r, r_{-i}) \leq D(r', r_{-i})$, for all $r' \in r$, and $r = BR(r_{-i})$.

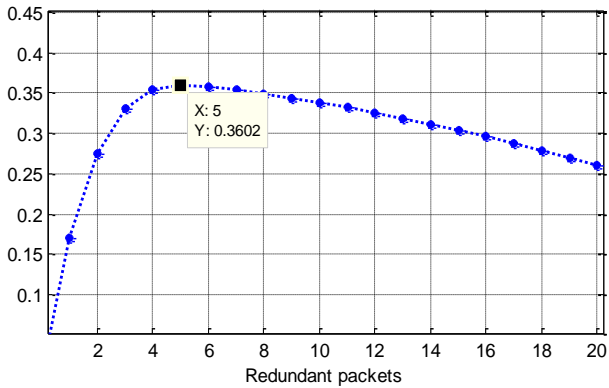

Figure 3.  The network Throughput vs r.

*E.  Discussion of results*

The impact of working at equilibrium point is clear in Fig. 5 that illustrates a comparison between three different modes of N-p CSMA; Conventional network (blue line), network with MDS EC out of equilibrium (red line) and network with MDS EC at Nash equilibrium (green line). At equilibrium users transmit with lower transmission delay, without the need to change their strategies ($r^*$). The transmission throughput will be then higher than before. In conventional mode; the network do not practice any technique of errors correction nor a technique for managing conflict situations like game theory that's why the its propagation delay is longer than in the other modes.

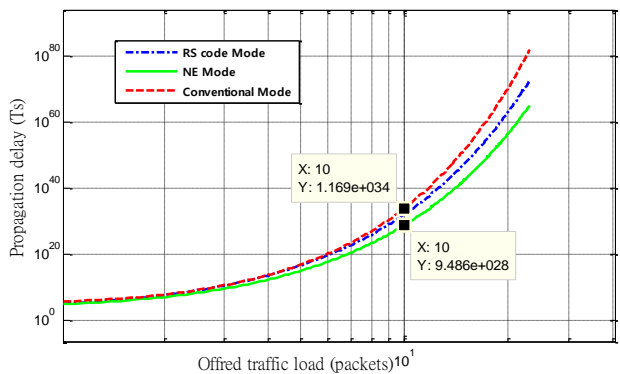

Figure 4.  The overall delay of CSMA network: • (red line: conventional (without EC). • (blue line: random MDS EC (out of NE)). • (green line: at equilibrium (NE))

In equilibrium mode, the network presents a smaller propagation delay than the other operating modes which explain the impact of choosing a suitable redundancy for MDS EC that's makes users satisfied about their outcomes or utility which is the overall propagation delay in our model. Here, the network is becoming more stable. Knowing that the axis of delay (vertical axis) is in logarithm scale and the difference then is very important (for ten packets, we have a big difference: $1,169.1034 - 9,486.1028 = 1.16.1034$ Ts. Thus, the advantage of making the network works at equilibrium may be more important with big data traffic (traffic load) or with big number of users.

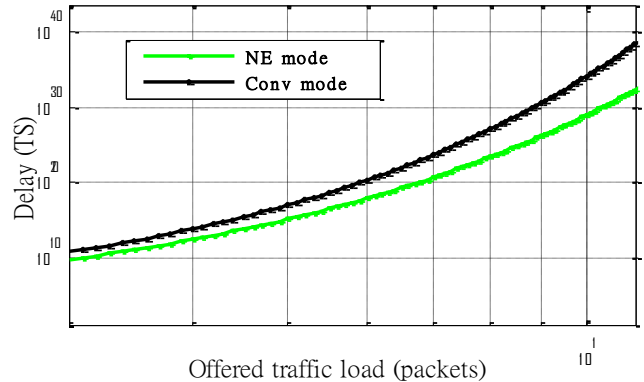

Figure 5.  Delay comparison: Equilibrium vs. Conventional network.

Fig. 5 illustrates better the advantage of equilibrium when comparing the propagation delay for network at equilibrium with that in out of equilibrium. The advantage (the difference) may be bigger if we increase the traffic load (traffic load > 10 packets), taking into consideration the risk of going to saturation mode in case of many or repeated collisions. Fig. 6 also, explains more clearly the advantage of equilibrium point.

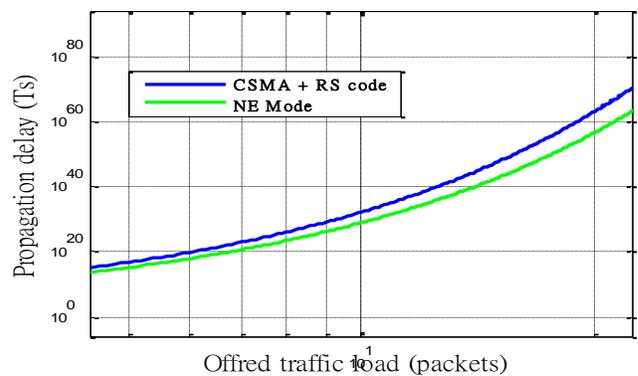

Figure 6.  Delay comparison: Equilibrium vs. Out of equilibrium.

## VI.  CONCLUSION

A non-cooperative game in N-p CSMA network is presented in this paper. The MDS EC is integrated to recover collided and erased packets to improve the transmission reliability. MDS EC should be applied properly. We presented a game model based on game theory to well manage the network und guide it to be more stable and efficiency. The advantage of game theory in CSMA protocols is clearly demonstrated and verified by appropriate simulations. Where, at equilibrium the overall propagation delay is becoming smaller than that in out of equilibrium. At equilibrium, the network can reach the minimum level of propagation delay with big difference comparing with conventional mode and that in out of equilibrium mode. In equilibrium, the network operate in development mode far from saturation which make it more stable with lower propagation delay and consequently with higher network throughput.

## VII. FUTURE WORK

Game theory offers many techniques and scenarios to solve different issues in wireless networks. Simulation results confirm that the proposed model improves the quality of network via minimizing the overall propagation delay when operating at the NE. Nevertheless, the effect of the occasional change in a minimum distance on the coding process remains to be studied and verified. So, research development should focus on coding process with the recurrent variation of redundancy. The next step of this research should be involved on the study of a variable MDS EC versus the frequent variation of redundancy in order to ensure the feasibility and good behaviour of the proposed model.